\documentclass[%
 reprint,
showpacs,%preprintnumbers,
 amsmath,amssymb,
 aps,
prl,
]{revtex4-1}

\usepackage[all,matrix,arrow,curve]{xy}
\def\V{\mathbb{V}}
\def\W{\mathbb{W}}

\usepackage[pdftex]{graphicx}% Include figure files
\usepackage{dcolumn}% Align table columns on decimal point
\usepackage{bm}% bold math
\def\be{\begin{equation}}
\def\ee{\end{equation}}
\def\ba{\begin{eqnarray}}
\def\ea{\end{eqnarray}}

\def\rd{\mathrm{d}}

\def\g{\mathfrak{g}}

\let\a=\alpha \let\b=\beta

\newcommand*{\R}{{\mathbb R}}

\begin{document}

\title{Non-abelian Gerbes: \\ Most general invariant action functional in generic dimensions}

\title{Non-abelian Gerbes and Semi-Courant-Dorfman Algebras}

\title{Non-abelian Gerbes and Enhanced Leibniz Algebras}

\author{Thomas Strobl}\email{strobl@math.univ-lyon1.fr}
\affiliation{Institut Camille Jordan,
Universit\'e Claude Bernard Lyon 1 \\
43 boulevard du 11 novembre 1918, 69622 Villeurbanne cedex,
France
}%
\date{April 6, 2016}
\begin{abstract}
We present the most general gauge-invariant action functional for coupled 1- and 2-form gauge fields with kinetic terms in generic dimensions, i.e.~dropping eventual contributions that can be added in particular space-time dimensions only such as higher Chern-Simons terms. After appropriate field redefinitions it coincides with a truncation of the Samtleben-Szegin-Wimmer action. In the process one sees explicitly how the existence of a gauge invariant functional enforces that the most general semi-strict Lie 2-algebra describing the bundle of a non-abelian gerbe gets reduced to a very particular structure, which, after the field redefinition, can be identified with the one of an enhanced Leibniz algebra. This is the first step towards a systematic construction of such functionals for higher gauge theories, with kinetic terms for a tower of gauge fields up to some highest form degree $p$, solved here for $p=2$. 
\end{abstract}

\maketitle

\section{Introduction}
While the construction of the field content, the gauge transformations, and the corresponding bundles for higher gauge theories have been much developed over the last decade, by different authors and different approaches \cite{ACJ05,Aschieri:2004yz,BM05,Baez02jn,BS05,Wockel,SchreiberStasheff,Wagemann,Waldorf,Saemann:2012uq,Jurco:2014mva}, gauge invariant functionals are rare. There are a few exceptions to this \cite{Henneaux-Knaepen,Henning1,Henning2,Generalizing,CYMH}, but, up to now, a general framework and in particular one that is simultaneously systematic and limiting the technical difficulties to a resonable amount of calculations, definitely has been missing so far. It is the intention of the present short communication to make a first step in filling this gap.

To set up a general procedure for arbitrary highest form degree $p$ for  a tower of gauge fields starting with 1-forms or even 0-forms, a super-geometric approach, as the one developed in  \cite{Gruetzmann-Strobl,KS07} and applied in \cite{Lavau}, will be indispensable: The problem  reduces to an equivariant extension of a super vielbein together with a vector field, both defined on a graded manifold that is associated to the higher gauge theory from the start on. While pointing out some features that hold for any $p$ in this note, we will mainly restrict our attention to the transition from $p=1$, the case of pure Yang-Mills gauge theories, to $p=2$, describing non-abelian gerbes. Therefore here we will mostly avoid the super language so as to not get burried in the formalism, and discuss instead the main changes that occur already in this first extension in a predominantly gauge theoretic terminology.

\section{The general gauge structure}

In the following we recall the main formulas from \cite{Gruetzmann-Strobl} needed for the sequel. For a higher gauge theory with form degrees up to $p$, we introduce a positively graded manifold ${\cal{}M}$ where each independent gauge field of form degree $k$ corresponds to one coordinate of the same parity as the differential form. For example, for $p=2$, one has a couple of coordinates $(\xi^a)_{a=1}^r$ of degree 1 and thus anti-commuting and $(b^I)_{I=1}^s$ of degree 2 and commuting; this corresponds to $r$ 1-form gauge fields $A^a$ and $s$ 2-form gauge fields $B^I$ interacting among one another. Then we write the most general degree plus one vector field $Q$ on ${\cal{}M}$ and require it to square to zero. In this way we introduce the gauge structure of the theory. For the case of our main interest, $p=2$,
\begin{eqnarray}\label{Q}
Q&=&\left(-\frac{1}{2} C^c_{ab} \xi^a \xi^b +  t_I^c b^I \right) \frac{\partial}{\partial \xi^c} +\\ &&+ \left( -\alpha^J_{Ia} b^I \xi^a + \frac{1}{6}\gamma^J_{abc}\xi^a\xi^b\xi^c\right) \frac{\partial}{\partial b^J}\nonumber \, ,
\end{eqnarray}
and $Q^2=0$ encodes the identities to be satisfied by the structure constants $C^c_{ab}$, $t_I^c$, $\alpha^J_{Ia}$, $\gamma^J_{abc}$. These in total \emph{five} families of identities are the adequate generalization of the Jacobi identities for the usual structure constants $C^c_{ab}$. For the example \eqref{Q}, this parametrizes a (semi-strict) Lie 2-algebra \cite{Baez03vi}, while for general $p$ one obtains a Lie p-algebra, i.e.~a particular type of $L_\infty$-algebra \cite{Lieinfinityalg}. A graded manifold equipped with such a vector field $Q$ is called a Q-manifold \cite{Schwarz,AKSZ}. According to \cite{Gruetzmann-Strobl}, it contains all the necessary general gauge structure of a higher form degree gauge theory, independent of space-time dimensions or particularities of special functionals. 

Let us denote the gauge fields of all different form degrees collectively by $A^\alpha$. Then the corresponding field strengths take the form
\begin{equation}\label{Falpha}
F^\alpha = \rd A^\alpha - Q^\alpha(A)
\end{equation}
where $Q^\a(A)$ denotes the $\alpha$-component of the vector field $Q$, subsequently replacing all graded coordinates in this polynomial by the corresponding gauge fields, with the multiplication being the wedge-product. For our preferred example $p=2$, this yields
\begin{eqnarray}
 \label{Fa}
F^a & =& \rd A^a+ \frac{1}{2} C^c_{ab} A^a \wedge A^b -t_I^c B^I \, , \\
\label{FI} F^I &=& \rd B^I + \alpha^I_{aJ} A^a \wedge B^J - \frac{1}{6} \gamma^I_{abc} A^a \wedge A^b \wedge A^c \, .
\end{eqnarray}
Consider the locally defined algebra $\cal{}A$ generated by the differential forms $A^\alpha$ and $\rd{}A^\alpha$. Sums of ``words'' in these ``letters'' that contain at least one $F^\alpha$ form an ideal ${\cal{}I}\subset{\cal{}A}$. One still remains inside the same ideal when performing redefinitions of its generators such as 
\begin{equation} \label{GI}
G^I = F^I - \beta^I_{a b} A^a \wedge F^b 
\end{equation} 
for some at this point arbitrary choice of constants $\beta^I_{ab}$---and likewise so for other field strengths, using those appearing already before within the tower. Only the first one, $F^a$, remains unmodified by such redefinitions.

Certainly, in particular space-time dimensions one could add further terms. For example, for space-time dimension $d=4$, one can add a term such as $\nu^I_{ab}A^a\wedge{}*F^b$ to \eqref{GI}. The considerations of the present paper are for generic dimensions, or the structure that is common to all dimensions $d$, and then \eqref{GI} describes the most general redefinition or extension of the system \eqref{Fa}, \eqref{FI}.\footnote{The basis  $F^\alpha$ of generators of ${\cal{}I}$ is uniquely singled out by the property that the only term containing a de Rham differential is the first one, $\rd{}A^\alpha$. However, for squaring inside an action functional, as needed in  a kinetic term for $A^\alpha$, such a choice of field strengths will turn out to be too restrictive.}

The generalized version of Bianchi identities becomes very elegant in this formulation: $\rd{\cal{}I}\subset{\cal{}I}$. In fact, this condition leads one necessarily to the $Q$-structure behind any higher gauge theory \cite{Gruetzmann-Strobl}. 

Infinitesimal gauge transformations are parametrized by some $\varepsilon^\alpha$, one for each $A^\alpha$ and of form degree one less than the one of the corresponding gauge field. As generators for the gauge transformations, stability of ${\cal{}I}$ leads one to consider
\begin{equation}\label{delta0}
\delta^0 A^\a = \rd \varepsilon^\a + \varepsilon^\beta (\partial_\beta Q^\alpha)(A) \, , 
\end{equation}
to which one may again add some $F$-dependent new terms, $\delta{}A^\a{}=\delta^0A^\a+\delta^1A^\a$ where $\delta^1A^\a\subset{\cal{}I}$. These transformations have the characteristic property
that $\delta{\cal I}\subset{\cal I}$;  even more,  $\delta^0{\cal I}\subset{\cal I}$ does not only determine the ``main part'' $\delta^0$ of the symmetries to be of the form \eqref{delta0} for the same $Q$ as before,  for given ${\cal I}$ this condition can be used to  find simultaneously the form of $F^\alpha$ and $\delta^0 A^\a$ as well as the main structural property $Q^2=0$ \cite{Gruetzmann-Strobl}. For a transformation of $F^\alpha$ with respect to the main part of the symmetries \eqref{delta0}, an explicit formula is known:
\begin{equation} \label{deltaF0}
\delta^0 F^\alpha = (-1)^{|\beta|+1} \varepsilon^\beta F^\gamma (\partial_\gamma \partial_\beta Q^\alpha)(A)\, ,
\end{equation}
where $|\beta|$ denotes the form degree of $A^\beta$.
Let us now specialize the last two paragraphs for $p=2$ (or a general $p$, but for the first two levels of the tower): 
\begin{eqnarray}
\delta^0 A^a &=& \rd \varepsilon^a + C^a_{bc} A^b \varepsilon^c + t^a_I \mu^I \\
\delta^0 B^I  &=& \rd \mu^I  + \alpha^I_{aJ} (A^a  \mu^J - \varepsilon^a B^I) + \frac{1}{2}\gamma^I_{abc} A^a  A^b \varepsilon^c \\
\delta^0 F^a &=& -C^a_{bc} \varepsilon^b F^c \label{delta0Fa} \\
\delta^0 F^I &=& -\alpha^I_{aJ} \varepsilon^a F^J + 
(\alpha^I_{aJ} \mu^J - \gamma^I_{abc}\varepsilon^b A^c) F^a \label{delta0FI}
\end{eqnarray}
where wedge products are understood. While necessarily
$\delta A^a\equiv\delta^0A^a$, one has $\delta B^I\equiv\delta^0B^I+\delta^1B^I$ for some $\delta^1 B^I = \sigma^I_{ab} \varepsilon^a F^b$. Anticipating considerations from below---in particularly, that one necessarily needs a transformation law of $G^I$ in \eqref{GI} of the form of Eq.~\eqref{Galpha} below---we will require that the transformation of $G^I$ in \eqref{GI} does not pick up a term proportional to $\rd \varepsilon$. Then a simple exercise yields $\sigma^I_{ab} = - \b^I_{ab}$, and thus the complete gauge symmetries for $p=2$ take the form
\begin{equation} \label{delta}
\delta A^a = \delta^0 A^a \quad , \qquad 
\delta B^I = \delta^0 B^I +\beta^I_{ab} \varepsilon^a F^b \, .
\end{equation}
It is one of the main advantages of the present approach that the only further ambiguities in the construction of an action functional lie in the addition of $F$-terms of lower degrees. For $p=2$ this amounts thus to  the introduction of merely \emph{one} hitherto  undetermined new set of constants $\beta^I_{ab}$ in \eqref{GI} and \eqref{delta}.

\section{On the Construction of actions}
Following the pattern of \cite{LAYM}, generalized to arbitrary higher gauge theories in \cite{Generalizing}, a gauge invariant action functional without any further constraints on the structure constants except for the existence of an appropriately invariant metric $\kappa$, can be obtained by squaring the highest field strength and adding the lower ones by means of Lagrange multipliers. For $p=2$,
\begin{equation}
S[A^a,B^I,\Lambda_a] : = \int_\Sigma \Lambda_a \wedge F^a + \frac{1}{2}\kappa_{IJ} F^I \wedge * F^J \, ,
\end{equation}
where $\Lambda_a$ are $d\!-\!1$-forms on $d$-dimensional space-time $\Sigma$ and, according to \eqref{delta0FI}, the metric coefficients $\kappa$ have to satisfy $\alpha_{a(I}^L \kappa_{J)L}=0$. The goal of the present article, however, is to consider a gauge-consistent deformation of 
\begin{equation}
S_0[A^\alpha] = \frac{1}{2}\int_\Sigma \kappa_{\a \b} \rd A^\alpha \wedge *\rd A^\beta \, ,
\end{equation}
i.e.~a functional with kinetic terms for \emph{all} the gauge fields $A^\alpha$, extending $S_0$ in a way consistent with the higher gauge structure specified above. Here the most direct attempt \cite{Baez02jn}, $\int_\Sigma \kappa_{\a \b} F^\alpha \wedge *F^\beta$, turns out to be far too restrictive: As obvious from \eqref{delta0FI}, this would require $\alpha^I_{aJ}=0=\gamma_{abc}^I$ for $p=2$, reducing $F^I$ to the abelian $F^I=\rd B^I$. 

The way out here is to consider  actions of the form $S_{higher}[A^\alpha]=\frac{1}{2}\int_\Sigma \kappa_{\a \b} G^\alpha \wedge *G^\beta$ with correspondingly adapted gauge transformations $\delta A^\alpha = \delta^0 A^\alpha + \delta^1 A^\alpha$, in generalization of  \eqref{GI} and \eqref{delta}, where for higher $p$ additions can and should contain also higher wedge products of lower $F$s. 

Before solving this problem in full generality for $p=2$, let us draw some direct conclusions for general $p$. The variation of $S_{higher}$ can vanish only if for each fixed degree of $\alpha$, the $G^\alpha$s transform into themselves. This implies that necessarily there exist constants $c^\a_{\b a}$ such that 
\begin{equation}\label{Galpha}
\delta G^\a = -c^\a_{a\b }  \varepsilon^a G^\b\, .
\end{equation}    
In particular, all the $G$s need to be strictly \emph{invariant} with respect to $\varepsilon^\alpha$ with $|\a|\ge 2$. Since, moreover, the gauge transformations of a functional need to close on-shell always \cite{Henneaux-Teitelboim}, the field equations contain a Hodge dulatity operator, but the commutator of gauge transformations do not, the gauge transformations $\delta_\varepsilon$ leaving invariant $S_{higher}$ need to close strictly off-shell. Finally, the conditions to be satisfied by the non-degenerate scalar products $\kappa_{\alpha \beta}$ follows from \eqref{Galpha} to always be of the form \begin{equation} \label{ckappa}
c^\gamma_{a(\a } \kappa_{\beta) \gamma} = 0 \, ,
\end{equation}
for all the different form degrees of $A^\alpha$.

\section{Most general action for $p=2$}
Although $G^a \equiv F^a$, the gauge transformation of $F^a$ is no more of the form \eqref{delta0Fa}, but the coefficients $C^a_{bc}$ appearing there change to 
\begin{equation} c^a_{bc} = C^a_{bc} + t^a_I \beta^I_{bc} \, ,  \label{c}
\end{equation}  which now is no more necessarily anti-symmetric in its lower indices. Thus, the transition of the standard Yang-Mills case $p=1$ to the non-abelian gerbe theory $p=2$ does not only relax the condition of the Jacobi identity for the structure constants $C^a_{bc}$, in addition, according to \eqref{ckappa}, the metric $\kappa_{ab}$ of the 1-form gauge fields has to satisfy an invariance condition with respect to the \emph{modified} coefficients \eqref{c}: $c^d_{a(b}\kappa_{c)d}=0$. 

It now remains to calculate $\delta G^I$, which is somewhat lengthy but straightforward. Here one  also needs  the Bianchi identity $\rd F^a = -C^a_{bc} A^b F^c - t^a_I F^I$. As a result, one finds that the $\mu$-invariance of $G^I$ and its $\varepsilon$-equivariance \eqref{Galpha} 
hold true, iff
\begin{eqnarray}
\alpha^I_{aJ} &=& \beta^I_{ba} t^b_J \, , \label{one}\\
\gamma^I_{abc} &=&C_{ab}^d\beta^I_{dc}  -2C^d_{c[a} \beta^I_{b]d}  \label{two} \\ && 
- 2(\beta^I_{d[a} \beta^J_{b]c} +\beta^I_{[a|d} \beta^J_{b]c})t^d_J \, ,\nonumber 
%\\
%\gamma^I_{abc} &=&-C_{ab}^d\beta^I_{dc}  +2C^d_{c[a} \beta^I_{b]d}  
%- 2\beta^I_{(d[a)} \beta^J_{b]c}t^d_J
\end{eqnarray}
respectively. In addition, one finds $c^I_{Ja} = \a^I_{\a J} + \beta_{ab}^I t^b_J = 2\beta_{(ab)}^I t^b_J $, which determines the condition of $\kappa_{IJ}$ for full gauge invariance of the action functional describing a non-abelian gerbe,
\begin{equation}\label{gerbe}
S_{gerbe} = \frac{1}{2} \int_\Sigma \kappa_{ab} F^a \wedge *F^b + \kappa_{IJ}G^I \wedge *G^J \, .
\end{equation}
We finally note that while $[\delta_\varepsilon,\delta_{\bar{\varepsilon}}]$ needs to close on $A^a$ for form degree reasons already, its closure on the 2-form fields $B^I$ requires one structural identity to hold true, which, however, can be seen to follow from  the two conditions \eqref{one} and \eqref{two} above. The composed parameters $\widetilde{\varepsilon}^\alpha$ remain unchanged by the additions $\delta^1$ guaranteeing closure \cite{Gruetzmann-Strobl}: $\widetilde{\varepsilon}^a = C^a_{bc} \varepsilon^b \bar{\varepsilon^c}$ and 
$\widetilde{\mu}^I = \alpha^I_{aJ} (\varepsilon^a \bar{\mu}^J - \bar{\varepsilon}^a \mu^J) - \gamma^I_{abc}  \varepsilon^a \bar{\varepsilon^b} A^c$. 
 
%\vspace{5mm}
\section{The gerbe Lie 2-algebra}
A good part of the present work goes into the identification of the mathematical structure underlying the gerbe action. Here we present the result of this analysis only: Corresponding to the $r$-dimensional internal space $\mathbb{V}$ of the 1-form gauge fields $A$ and the $s$-dimensional internal space $\mathbb{W}$ of the 2-form gauge fields $B$, one has the following short sequence 
\begin{equation}\label{VW}
\mathbb{W} \stackrel{t}{\longrightarrow} \mathbb{V} \, .
\end{equation}
together with a Leibniz product on $\mathbb{V}$, denoted by a  bracket,  $[ \cdot, \cdot ] \colon \mathbb{V} \otimes \mathbb{V} \to \mathbb{V}$, and satisfying, by definition, \begin{equation}\label{Leibniz}
[v_1,[v_2,v_3]]=[[v_1,v_2],v_3]+[v_2,[v_1,v_3]]
\end{equation}
for all $v_i \in \mathbb{V}$, 
and a product  $\circ$ on $\mathbb{V}$ with values in $\mathbb{W}$. These satisfy the following compatibilities: 
\begin{equation} \label{axioms}
t(w) \circ t(w) = 0 \: , \quad [v,v]=t(v\circ v) \: , \quad 
[t(w),v] = 0 \, ,
\end{equation}
for all $v \in \mathbb{V}$, $w \in \mathbb{W}$ as well as 
\begin{equation}\label{coherence}
u \stackrel{s}{\circ} [v,v] = v\stackrel{s}{\circ} [u,v]  
\end{equation}
for all $u,v \in \mathbb{V}$, where $\stackrel{s}{\circ}$ denotes the symmetrization of the product $\circ$, i.e.~$u \stackrel{s}{\circ} v \equiv \frac{1}{2} (u \circ v + v \circ u)$. After a choice of basis $e_a$ in $\mathbb{V}$ and $b_I$ in $\mathbb{W}$, the identification with the components is given by:
\begin{equation}
t(b_I) = t^a_I e_a \: , \quad  [e_a,e_b] = c^c_{ab} e_c \: , \quad e_a \circ e_b = \beta^I_{ab} b_I \, .
\end{equation}
Remarkably, all the eight, in part involved equations following from the previous considerations are satisfied identically with these axioms. We remark here only that \eqref{two} does not just determine $\gamma^I_{abc}$ in terms of other quantities, but also leads directly to the constraint \eqref{coherence} due to its total skew-symmetry. 
 
In particular, by construction, the above structure gives rise to a semi-strict Lie 2-algebra, which, 
as a $\gamma$-twisted crossed module of Lie algebras, is obtained from the following identifications: For the bracket on  $\mathbb{V}$ one takes  $[u,v]_{\mathbb{V}} := [u,v]-t(u \circ v)$, for the one on $\mathbb{W}$, $[w_1,w_2]_{\mathbb{W}} = -t(w_1) \circ t(w_2)$,  both of which are antisymmetric due to the first two equations in \eqref{axioms}. The (twisted) action $\star$ of $\mathbb{V}$ on $\mathbb{W}$ is given by $v \star w := t(w) \circ v$,  while the formula for the ``anomaly'' $\gamma$ is somewhat longer:
\begin{eqnarray}\label{gamma}
\gamma(v_1,v_2,v_3)& =& [v_1,v_2]\circ v_3 + v_2 \circ [v_1,v_3] - v_1 \circ [v_2,v_3] \nonumber \qquad {} \\ 
&&\! \!\! \!\! \! \! \!\! \!\! \! \! \!\! \!\! \! \! \!\! \!\! \! \! \!-t(v_1 \circ v_2) \circ v_3 + t(v_1\circ v_3) \circ v_2 - t(v_2 \circ v_3)\circ v_1. \!\!\!\!{}%\nonumber 
\end{eqnarray}
 If $\gamma$ vanishes, this becomes a strict Lie 2-algebra or crossed module: the brackets  $[\cdot,\cdot]_{\mathbb{V}}$ and $[\cdot,\cdot]_{\mathbb{W}}$ are then Lie brackets, $t$ a Lie algebra morphism,  and $\star$ becomes a map from $\mathbb{V}$ into the derivations of $\mathbb{W}$. 

It is one of the main findings of this paper that 
only \emph{those} (semi-strict) Lie 2-algebras which arise in the above way result in a gauge invariant action functional of the form $S_{gerbe}$.\footnote{Note that in \cite{Baez02jn} the important question of gauge invariance was ignored and the functional provided there was in fact not invariant.}

%The above is a particular semi-strict Lie 2-algebra, the Lie 2-algebra describing the gauge structure of the action $S_{gerbe}$. 

We conclude this section by remarking that the condition of the metric $\kappa_\mathbb{V}$ on $\mathbb{V}$ needed for gauge-invariance of $S_{gerbe}$ is simply  invariance with respect to left action of the Leibniz bracket, while  $\kappa_\mathbb{W}$ needs to be invariant with respect to the operator $v \stackrel{s}{\circ} t(\cdot) $ on $\mathbb{W}$ for any $v \in \mathbb{V}$.
%, i.e.~the $\star$-operation mentioned above.

%

\section{Reduced action and algebra}
The physics described by the functional $S_{gerbe}$ remains unchanged if we perform field redefinitions such as $B^I \mapsto B^I - \frac{1}{2}\beta_{ab}^I A^a \wedge A^b$. A simple look at \eqref{GI} and \eqref{FI} shows that this kills precisely the antisymmetric part of the coefficient $\beta^I_{ab}$ in the first of these two equations. Also other coefficients change as is best seen by performing the super-diffeomorphism 
\begin{equation}\label{b}
b^I \mapsto b^I- \frac{1}{2}\beta_{ab}^I \xi^a \wedge \xi^b
\end{equation}
within the vector field \eqref{Q}. But we do not need to follow all the changes that the other structure constants undergo, which can be found in \cite{Gruetzmann-Strobl};\footnote{Most dramatic is the simplification in \eqref{gamma}. As seen best from \eqref{two}, where the second line vanishes completely after anti-symmetrization in the three indices, one finds $\gamma(v_1,v_2,v_3)=\frac{1}{3}([v_1,v_2]_\V \circ v_3 + cycl)$.} after this change of fields and constants performed at the very beginning of the calculation, its net effect is simply that the product $\circ$ is symmetric now.

 With this simplification, the functional \eqref{gerbe} becomes \emph{identical} to a truncation of the theory found in \cite{Henning1}.\footnote{Identify for this $C^a_{bc}$ with $-f^a_{bc}$, $t^a_I$ with $-h^a_I$ and $\beta^I_{ab}$ with the negative of the symmetric $d^I_{ab}$ and put to zero the 3-form field together with $b_{Iab}$ and $g^{Ia}$.} We started out with a complete and systematic approach, the coincidence of the most general action for $p=2$ with the previously found ``example'' is therefore remarkable. 

We conclude the discussion of $p=2$ by describing its underlying algebraic structure: For a symmetric operation $\circ$, the axioms \eqref{Leibniz}, \eqref{axioms}, and \eqref{coherence} describe what can be called a semi Courant-Dorfman algebra \cite{Courant-Dorfman-algebras}. Courant-Dorfman algebras, which contain additional structure not present here, were introduced as a joint axiomatization of Courant algebroids \cite{Courant-algebroids,Severa-Weinstein} and string current algebras \cite{Alekseev-Strobl}. For the identification, one regards $\circ$ as a scalar product $(\cdot, \cdot)$ on $\V$ with values in $\W$. Replacing $t$ by $\frac{1}{2} \partial$, the condition \eqref{coherence} now obtains the interpretation of invariance of this scalar product with respect to the action of the Leibniz bracket on $\V$ and $\circ \partial$ on $\W$: $u \circ \partial (v_1,v_2) = ([u,v_1],v_2) + (v_1,[u,v_2])$. In particular, \emph{every} Courant-Dorfman algebra gives rise to a Lie 2-algebra (in generalization of the result for Courant algebroids \cite{Roytenberg-Weinstein}) and can serve as the gauge structure of a non-abelian gerbe.

More important than this observation is the fact that already \emph{every} Leibniz algebra structure $(\V,[\cdot,\cdot])$ gives rise to such a structure in the following way: Let $\W_0$ be the ideal generated by squares $[v,v]$ in $\V$ and $t_0$ its embedding into $\V$. One then verifies that \emph{all} the axioms \eqref{axioms} and \eqref{coherence} are automatically satisfied.\footnote{More details on this will be provided in \cite{Alexei-Henning-Strobl}.} Reciprocally, if $t \colon \W \to \V$ is injective and $\circ \colon S^2 \V \to \W$ is surjective, the complete structure \emph{is} of this form. In general, certainly the map $t$ can have a kernel and its image can be strictly larger than $\W_0$, while necessarily the image of the composition of $\circ$ with $t$ needs to coincide with $t$ due to the second equation in \eqref{axioms}. The general situation is depicted in Fig.~\ref{fig:Leibniz}.
\begin{figure}
\centering
\hskip0mm
\xymatrix{ 
 \ker t \:\ar@{^{(}->}[r] &  \W \ar[dr]^{t} \\ 0\ar[r] & \W_0
 \ar[r]^{t_0} & \V \ar@{->>}[dr] \ar@{->>}[r] 
 & 
 \mathfrak{g}_{0} \ar@{->>}[d]
\\ &&& \mathfrak{g}_t }
\caption{The vertical line shows the canonical data of any Leibniz algebra $\V$: $\W_0$ is the embedded ideal of squares and $\g_0$ the canonical quotient Lie algebra. The pair $t \colon \W \to \V$ can be an enlargement of this structure in two ways: First, the map $t$ does not need to be an embedding, i.e.~$t$ can have a kernel, and second, the image of $t$ can be larger than $\W_0 \subset \V$; in this case, there still exists a quotient Lie algebra $\g_t$, which is simultaneously a quotient of $\g_0$.}
\label{fig:Leibniz}
\end{figure}
Given a Leibniz algebra $\V$, the enhancement to $t \colon \V \to \W$ together with $\circ \colon S^2 \V \to \W$ is quite restricted. For example, only the part of $\circ$ mapping into $\ker t$ is not already fixed by \eqref{axioms}. 

Still, there is some freedom that remains to be chosen. To give a simple example, take $\V$ and $\W$ to be both $\R^2$, $t$ the identification of the first summands, and the Leibniz bracket to be identically zero. Then there is precisely a two-parameter family of choices for $\circ$: $(x,y) \circ (x',y') = (0, c_1 y y' + c_2 (x y' + x' y))$. For $c_1=1$ and $c_2 = 0$, the resulting gauge theory \eqref{gerbe} is particularly simple: $F^{1} = \rd A^1-B^1$, $F^{2} = \rd A^2$, $G^1=\rd B^1$ and $G^2=\rd B^2-A^2\wedge \rd A^2$. The physics of this simple model is seen to describe one abelian massless gerbe gauge field $B^2$ interacting with one abelian massless gauge field $A^2$ and, after eliminating $A^1$ by a shift of $B^1$, one more \emph{massive} abelian free gerbe $B^1$. We have not discussed such type of field redefinitions in general for the action \eqref{gerbe}; but evidently they should be taken into account as well for a proper understanding of the resulting physics.

A whole class of less trivial examples can be constructed by looking at vector fields of a Q-manifold $\widetilde{{\cal M}}$ equipped with $\widetilde{r}$ degree 1  and $\widetilde{s}$ degree 2 coordinates: Degree -1 vector fields $V$ and degree -2 vector fields $W$ on $\widetilde{{\cal M}}$ are identified with elements  $v \in \mathbb{V}$ and $w \in \mathbb{W}$, respectively, the 
Lie bracket $[V,V]$ with $v\circ v$, the derived bracket $[[V,Q],V']$ with the Leibniz bracket $[v,v']$, and $t \colon \mathbb{W} \to \mathbb{V}$ is induced by $\mathrm{ad}_Q$. Enhanced Leibniz algebras, i.e.~structures defined by \eqref{VW}, \eqref{Leibniz}, \eqref{axioms}, and \eqref{coherence}, obtained in this way always have a non-degenerate, surjective product $\circ$, in which case the axioms can be proven to be equivalent to a $\mathbb{W}$-twisted Courant algebroid \cite{Gruetzmann-Strobl}  over a point. 

Examples stemming from $\widetilde{{\cal M}}$ like this have the feature that necessarily $\dim\mathbb{V}\equiv{}r=\widetilde{r}(1+\widetilde{s})>\dim\mathbb{W}\equiv{}s = \widetilde{s}$. For the construction of other examples, it seems reasonable to start with the Leibniz algebra on $\V$. Let us take an example where $r=s = 4$: The following defines a Leibniz bracket on $\V$, $[u,v]=u_4 (v_1 e_1 - v_2 e_2)$. Evidently, the subspace $\W_0$ of squares is $\langle e_1, e_2 \rangle \cong \R^2$ in this case. For $t$ we choose $t(b_I) = e_I$ for $I=1,2,3$ and $t(b_4)=0$ and for the (symmetric) $\circ$-product, $v \circ v = v_4(v_1b_1 - v_2 b_2 + k v_4 b_4)$, where $k \in \{ \pm 1, 0 \}$. We thus obtained an enhancement of the Leibniz algebra $\V$ with both, a non-trivial kernel of $t$ and, for non-zero $k$, with $t_0(\W_0) \subsetneq t(\W) \subsetneq \V$. 

These data give rise to a \emph{strict} Lie 2-algebra, since $\gamma$ turns out to vanish. At the price of adding a skew-symmetric part to $\circ$, it can be transformed into a non-strict one by a super-diffeomorphism of the type \eqref{b}.
  
In general, the appropriately invariant metrics $\kappa_\V$ and $\kappa_\W$ are \emph{additional} structures and their existence, as for Lie algebras, can restrict the admissible enhanced Leibniz algebras. This is well illustrated by the above example: $\kappa_\V([u,v],v)=0$ for all $u,v \in \V$ \emph{cannot} be satisfied for a positive definite $\kappa_\V$ since it is easily seen to lead to $\kappa_\V(e_1,e_1)=0$. This fact is even independent of the enhancement here.

\section{Conclusion and Outlook}
In this paper we continued the systematic construction of higher gauge theories started in \cite{Gruetzmann-Strobl} with gauge invariant functionals. For the purpose of a first orientation we focused on the case $p=2$. We showed that the most general ansatz in generic dimensions with kinetic terms for the 1-form and the 2-form gauge fields leads necessarily to a truncation of the Samtleben-Szegin-Wimmer action, at least after a field redefinition. Note in this context that while the iterative construction in \cite{Henning1,Henning2} includes within $G^\alpha$  at each step already the gauge field of degree $|\alpha|+1$, we found it advantageous to avoid this here: without it, the iteration $p \to p+1$ follows the tower of Lie $p$-algebras, which otherwise is obtained only after further truncation or projection \cite{Lavau}.

We furthermore observed that the Lie algebra structure on $\V$ for $p=1$ is turned into a Leibniz algebra upon transition to $p=2$. In fact, as will be shown in \cite{Alexei-Henning-Strobl}, any Leibniz algebra gives rise to a canonical Lie$_\infty$ algebra and it is this algebraic structure that underlies the tensor hierarchy of \cite{tensor1,tensor2,tensor3}. In \cite{Henning1}, on the other hand, the gauge structure is not determined completely by this Leibniz algebra, but by an appropriate enhancement of it, summarized in  Fig.~\ref{fig:Leibniz}.

It is clear from the present considerations that extending the construction to higher $p$ or also to ``just'' including scalar fields into the tower, will lead to an explosion of technical work. While for higher and higher $p$, the additional data will correspond to more and more involved algebraic structures, the inclusion of scalar fields turns this into a more and more involved problem within differential geometry \cite{Gruetzmann-Strobl}, \cite{CYMH}. All these complicated structures, however, promise to have a simple super-geometrical interpretation. We found that it is not the right strategy to focus on the closure of the gauge symmetries, since already for $p=2$ this yields only part of the necessary restrictions of the structure functions, but to instead turn to the extension problem of $G^\alpha$ satisfying the equivariance condition \eqref{Galpha}. With \eqref{Falpha}, \eqref{delta0}, and \eqref{deltaF0} we have closed formulas for any $p$. In the present approach, all the ambiguity lies in adding terms lying within the ideal $\cal I$. We intend to come back to a development of this idea in a sequel to this work. 

Using the approach \cite{KS07},  action functionals like \eqref{gerbe} will extend to a functional on non-trivial bundles. It is an interesting open problem for the future already for $p=2$ to relate the picture of $Q$-bundles to the alternative one of categorified bundles and bundle gerbes \cite{ACJ05,Wockel,Jurco:2014mva,SchreiberStasheff}.

%\newpage
\begin{acknowledgments}
It is a pleasure to thank H.~Samtleben for numerous discussions on tensor hierarchies and an important remark on a preliminary version of this paper. 
I gratefully acknowledge stimulating discussions with C.~Saemann about non-abelian gerbes as well as discussions with A.~Kotov, S.~Lavau and F.~Wagemann, related directly or indirectly to the present work. I further want to thank S.~Watamura and the Tohoku Forum for Creativity for their invitation to a very lively workshop from which I also profited by its inspiration to this work. This research was partially supported by Projeto P.V.E. 88881.030367/
2013-01 (CAPES/Brazil), A.~Alekseev's project MODFLAT of the European Research Council
(ERC), and the NCCR SwissMAP of the Swiss National Science Foundation. I am grateful for this support.
\end{acknowledgments}

%\bibliography{bibtexPRL}

\begin{thebibliography}{39}%
\makeatletter
\providecommand \@ifxundefined [1]{%
 \@ifx{#1\undefined}
}%
\providecommand \@ifnum [1]{%
 \ifnum #1\expandafter \@firstoftwo
 \else \expandafter \@secondoftwo
 \fi
}%
\providecommand \@ifx [1]{%
 \ifx #1\expandafter \@firstoftwo
 \else \expandafter \@secondoftwo
 \fi
}%
\providecommand \natexlab [1]{#1}%
\providecommand \enquote  [1]{``#1''}%
\providecommand \bibnamefont  [1]{#1}%
\providecommand \bibfnamefont [1]{#1}%
\providecommand \citenamefont [1]{#1}%
\providecommand \href@noop [0]{\@secondoftwo}%
\providecommand \href [0]{\begingroup \@sanitize@url \@href}%
\providecommand \@href[1]{\@@startlink{#1}\@@href}%
\providecommand \@@href[1]{\endgroup#1\@@endlink}%
\providecommand \@sanitize@url [0]{\catcode `\\12\catcode `\$12\catcode
  `\&12\catcode `\#12\catcode `\^12\catcode `\_12\catcode `\%12\relax}%
\providecommand \@@startlink[1]{}%
\providecommand \@@endlink[0]{}%
\providecommand \url  [0]{\begingroup\@sanitize@url \@url }%
\providecommand \@url [1]{\endgroup\@href {#1}{\urlprefix }}%
\providecommand \urlprefix  [0]{URL }%
\providecommand \Eprint [0]{\href }%
\providecommand \doibase [0]{http://dx.doi.org/}%
\providecommand \selectlanguage [0]{\@gobble}%
\providecommand \bibinfo  [0]{\@secondoftwo}%
\providecommand \bibfield  [0]{\@secondoftwo}%
\providecommand \translation [1]{[#1]}%
\providecommand \BibitemOpen [0]{}%
\providecommand \bibitemStop [0]{}%
\providecommand \bibitemNoStop [0]{.\EOS\space}%
\providecommand \EOS [0]{\spacefactor3000\relax}%
\providecommand \BibitemShut  [1]{\csname bibitem#1\endcsname}%
\let\auto@bib@innerbib\@empty
%</preamble>
\bibitem [{\citenamefont {Aschieri}\ \emph {et~al.}(2005)\citenamefont
  {Aschieri}, \citenamefont {Cantini},\ and\ \citenamefont {Jurco}}]{ACJ05}%
  \BibitemOpen
  \bibfield  {author} {\bibinfo {author} {\bibfnamefont {P.}~\bibnamefont
  {Aschieri}}, \bibinfo {author} {\bibfnamefont {L.}~\bibnamefont {Cantini}}, \
  and\ \bibinfo {author} {\bibfnamefont {B.}~\bibnamefont {Jurco}},\
  }\href@noop {} {\bibfield  {journal} {\bibinfo  {journal} {Commun. Math.
  Phys.}\ }\textbf {\bibinfo {volume} {254}},\ \bibinfo {pages} {367} (\bibinfo
  {year} {2005})},\ \Eprint {http://arxiv.org/abs/hep-th/0312154}
  {hep-th/0312154} \BibitemShut {NoStop}%
\bibitem [{\citenamefont {Aschieri}\ and\ \citenamefont
  {Jurco}(2004)}]{Aschieri:2004yz}%
  \BibitemOpen
  \bibfield  {author} {\bibinfo {author} {\bibfnamefont {P.}~\bibnamefont
  {Aschieri}}\ and\ \bibinfo {author} {\bibfnamefont {B.}~\bibnamefont
  {Jurco}},\ }\href {\doibase 10.1088/1126-6708/2004/10/068} {\bibfield
  {journal} {\bibinfo  {journal} {JHEP}\ }\textbf {\bibinfo {volume} {10}},\
  \bibinfo {pages} {068} (\bibinfo {year} {2004})},\ \Eprint
  {http://arxiv.org/abs/hep-th/0409200} {arXiv:hep-th/0409200 [hep-th]}
  \BibitemShut {NoStop}%
%%CITATION = HEP-TH/0409200;%%
\bibitem [{\citenamefont {Breen}\ and\ \citenamefont {Messing}(2005)}]{BM05}%
  \BibitemOpen
  \bibfield  {author} {\bibinfo {author} {\bibfnamefont {L.}~\bibnamefont
  {Breen}}\ and\ \bibinfo {author} {\bibfnamefont {W.}~\bibnamefont
  {Messing}},\ }\href@noop {} {\bibfield  {journal} {\bibinfo  {journal} {Adv.\
  Math.}\ }\textbf {\bibinfo {volume} {198}},\ \bibinfo {pages} {76} (\bibinfo
  {year} {2005})},\ \Eprint {http://arxiv.org/abs/math/0106083} {math/0106083}
  \BibitemShut {NoStop}%
\bibitem [{\citenamefont {Baez}(2002)}]{Baez02jn}%
  \BibitemOpen
  \bibfield  {author} {\bibinfo {author} {\bibfnamefont {J.~C.}\ \bibnamefont
  {Baez}},\ }\href@noop {} {\  (\bibinfo {year} {2002})},\ \Eprint
  {http://arxiv.org/abs/hep-th/0206130} {hep-th/0206130} \BibitemShut {NoStop}%
%%CITATION = HEP-TH 0206130;%%
\bibitem [{\citenamefont {Baez}\ and\ \citenamefont {Schreiber}(2007)}]{BS05}%
  \BibitemOpen
  \bibfield  {author} {\bibinfo {author} {\bibfnamefont {J.~C.}\ \bibnamefont
  {Baez}}\ and\ \bibinfo {author} {\bibfnamefont {U.}~\bibnamefont
  {Schreiber}},\ }in\ \href@noop {} {\emph {\bibinfo {booktitle} {Categories in
  Algebra, Geometry and Mathematical Physics}}},\ \bibinfo {series} {Contemp.
  Math.}, Vol.\ \bibinfo {volume} {431},\ \bibinfo {editor} {edited by\
  \bibinfo {editor} {\bibfnamefont {A.}~\bibnamefont {Davydov}} \emph
  {et~al.}}\ (\bibinfo  {publisher} {AMS},\ \bibinfo {year} {2007})\ pp.\
  \bibinfo {pages} {7--30},\ \Eprint {http://arxiv.org/abs/math/0511710}
  {math/0511710} \BibitemShut {NoStop}%
\bibitem [{\citenamefont {{Wockel}}(2008)}]{Wockel}%
  \BibitemOpen
  \bibfield  {author} {\bibinfo {author} {\bibfnamefont {C.}~\bibnamefont
  {{Wockel}}},\ }\href@noop {} {\bibfield  {journal} {\bibinfo  {journal}
  {ArXiv e-prints}\ } (\bibinfo {year} {2008})},\ \Eprint
  {http://arxiv.org/abs/0803.3692} {arXiv:0803.3692 [math.DG]} \BibitemShut
  {NoStop}%
\bibitem [{\citenamefont {Sati}\ \emph {et~al.}(2008)\citenamefont {Sati},
  \citenamefont {Schreiber},\ and\ \citenamefont
  {Stasheff}}]{SchreiberStasheff}%
  \BibitemOpen
  \bibfield  {author} {\bibinfo {author} {\bibfnamefont {H.}~\bibnamefont
  {Sati}}, \bibinfo {author} {\bibfnamefont {U.}~\bibnamefont {Schreiber}}, \
  and\ \bibinfo {author} {\bibfnamefont {J.}~\bibnamefont {Stasheff}},\ }in\
  \href {\doibase 10.1007/978-3-7643-8736-5_17} {\emph {\bibinfo {booktitle}
  {{Quantum Field Theory, B. Fauser, J. Tolksdorf and E. Zeidler, Eds.,
  Birkhauser (2009) 303-424}}}}\ (\bibinfo {year} {2008})\ \Eprint
  {http://arxiv.org/abs/0801.3480} {arXiv:0801.3480 [math.DG]} \BibitemShut
  {NoStop}%
%%CITATION = ARXIV:0801.3480;%%
\bibitem [{\citenamefont {{Abbaspour}}\ and\ \citenamefont
  {{Wagemann}}(2012)}]{Wagemann}%
  \BibitemOpen
  \bibfield  {author} {\bibinfo {author} {\bibfnamefont {H.}~\bibnamefont
  {{Abbaspour}}}\ and\ \bibinfo {author} {\bibfnamefont {F.}~\bibnamefont
  {{Wagemann}}},\ }\href@noop {} {\bibfield  {journal} {\bibinfo  {journal}
  {ArXiv e-prints}\ } (\bibinfo {year} {2012})},\ \Eprint
  {http://arxiv.org/abs/1202.2292} {arXiv:1202.2292 [math.AT]} \BibitemShut
  {NoStop}%
\bibitem [{\citenamefont {{Nikolaus}}\ and\ \citenamefont
  {{Waldorf}}(2011)}]{Waldorf}%
  \BibitemOpen
  \bibfield  {author} {\bibinfo {author} {\bibfnamefont {T.}~\bibnamefont
  {{Nikolaus}}}\ and\ \bibinfo {author} {\bibfnamefont {K.}~\bibnamefont
  {{Waldorf}}},\ }\href@noop {} {\bibfield  {journal} {\bibinfo  {journal}
  {ArXiv e-prints}\ } (\bibinfo {year} {2011})},\ \Eprint
  {http://arxiv.org/abs/1112.4702} {arXiv:1112.4702 [math.AT]} \BibitemShut
  {NoStop}%
\bibitem [{\citenamefont {Saemann}\ and\ \citenamefont
  {Wolf}(2014)}]{Saemann:2012uq}%
  \BibitemOpen
  \bibfield  {author} {\bibinfo {author} {\bibfnamefont {C.}~\bibnamefont
  {Saemann}}\ and\ \bibinfo {author} {\bibfnamefont {M.}~\bibnamefont {Wolf}},\
  }\href {\doibase 10.1007/s00220-014-2022-0} {\bibfield  {journal} {\bibinfo
  {journal} {Commun. Math. Phys.}\ }\textbf {\bibinfo {volume} {328}},\
  \bibinfo {pages} {527} (\bibinfo {year} {2014})},\ \Eprint
  {http://arxiv.org/abs/1205.3108} {arXiv:1205.3108 [hep-th]} \BibitemShut
  {NoStop}%
%%CITATION = ARXIV:1205.3108;%%
\bibitem [{\citenamefont {Jurco}\ \emph {et~al.}(2015)\citenamefont {Jurco},
  \citenamefont {Saemann},\ and\ \citenamefont {Wolf}}]{Jurco:2014mva}%
  \BibitemOpen
  \bibfield  {author} {\bibinfo {author} {\bibfnamefont {B.}~\bibnamefont
  {Jurco}}, \bibinfo {author} {\bibfnamefont {C.}~\bibnamefont {Saemann}}, \
  and\ \bibinfo {author} {\bibfnamefont {M.}~\bibnamefont {Wolf}},\ }\href
  {\doibase 10.1007/JHEP04(2015)087} {\bibfield  {journal} {\bibinfo  {journal}
  {JHEP}\ }\textbf {\bibinfo {volume} {04}},\ \bibinfo {pages} {087} (\bibinfo
  {year} {2015})},\ \Eprint {http://arxiv.org/abs/1403.7185} {arXiv:1403.7185
  [hep-th]} \BibitemShut {NoStop}%
%%CITATION = ARXIV:1403.7185;%%
\bibitem [{\citenamefont {Henneaux}\ and\ \citenamefont
  {Knaepen}(1997)}]{Henneaux-Knaepen}%
  \BibitemOpen
  \bibfield  {author} {\bibinfo {author} {\bibfnamefont {M.}~\bibnamefont
  {Henneaux}}\ and\ \bibinfo {author} {\bibfnamefont {B.}~\bibnamefont
  {Knaepen}},\ }\href {\doibase 10.1103/PhysRevD.56.R6076} {\bibfield
  {journal} {\bibinfo  {journal} {Phys. Rev.}\ }\textbf {\bibinfo {volume}
  {D56}},\ \bibinfo {pages} {6076} (\bibinfo {year} {1997})},\ \Eprint
  {http://arxiv.org/abs/hep-th/9706119} {arXiv:hep-th/9706119 [hep-th]}
  \BibitemShut {NoStop}%
%%CITATION = HEP-TH/9706119;%%
\bibitem [{\citenamefont {Samtleben}\ \emph {et~al.}(2011)\citenamefont
  {Samtleben}, \citenamefont {Sezgin},\ and\ \citenamefont
  {Wimmer}}]{Henning1}%
  \BibitemOpen
  \bibfield  {author} {\bibinfo {author} {\bibfnamefont {H.}~\bibnamefont
  {Samtleben}}, \bibinfo {author} {\bibfnamefont {E.}~\bibnamefont {Sezgin}}, \
  and\ \bibinfo {author} {\bibfnamefont {R.}~\bibnamefont {Wimmer}},\ }\href
  {\doibase 10.1007/JHEP12(2011)062} {\bibfield  {journal} {\bibinfo  {journal}
  {JHEP}\ }\textbf {\bibinfo {volume} {12}},\ \bibinfo {pages} {062} (\bibinfo
  {year} {2011})},\ \Eprint {http://arxiv.org/abs/1108.4060} {arXiv:1108.4060
  [hep-th]} \BibitemShut {NoStop}%
%%CITATION = ARXIV:1108.4060;%%
\bibitem [{\citenamefont {Samtleben}\ \emph {et~al.}(2013)\citenamefont
  {Samtleben}, \citenamefont {Sezgin},\ and\ \citenamefont
  {Wimmer}}]{Henning2}%
  \BibitemOpen
  \bibfield  {author} {\bibinfo {author} {\bibfnamefont {H.}~\bibnamefont
  {Samtleben}}, \bibinfo {author} {\bibfnamefont {E.}~\bibnamefont {Sezgin}}, \
  and\ \bibinfo {author} {\bibfnamefont {R.}~\bibnamefont {Wimmer}},\ }\href
  {\doibase 10.1007/JHEP03(2013)068} {\bibfield  {journal} {\bibinfo  {journal}
  {JHEP}\ }\textbf {\bibinfo {volume} {03}},\ \bibinfo {pages} {068} (\bibinfo
  {year} {2013})},\ \Eprint {http://arxiv.org/abs/1212.5199} {arXiv:1212.5199
  [hep-th]} \BibitemShut {NoStop}%
%%CITATION = ARXIV:1212.5199;%%
\bibitem [{\citenamefont {Kotov}\ and\ \citenamefont
  {Strobl}(2010)}]{Generalizing}%
  \BibitemOpen
  \bibfield  {author} {\bibinfo {author} {\bibfnamefont {A.}~\bibnamefont
  {Kotov}}\ and\ \bibinfo {author} {\bibfnamefont {T.}~\bibnamefont {Strobl}},\
  }in\ \href {\doibase 10.4171/079-1/7} {\emph {\bibinfo {booktitle} {Handbook
  of pseudo-{R}iemannian geometry and supersymmetry}}},\ \bibinfo {series}
  {IRMA Lect. Math. Theor. Phys.}, Vol.~\bibinfo {volume} {16}\ (\bibinfo
  {publisher} {Eur. Math. Soc., Z\"urich},\ \bibinfo {year} {2010})\ pp.\
  \bibinfo {pages} {209--262},\ \Eprint {http://arxiv.org/abs/1004.0632}
  {1004.0632 [math.PH]} \BibitemShut {NoStop}%
\bibitem [{\citenamefont {Kotov}\ and\ \citenamefont
  {Strobl}(2015{\natexlab{a}})}]{CYMH}%
  \BibitemOpen
  \bibfield  {author} {\bibinfo {author} {\bibfnamefont {A.}~\bibnamefont
  {Kotov}}\ and\ \bibinfo {author} {\bibfnamefont {T.}~\bibnamefont {Strobl}},\
  }\href {\doibase 10.1103/PhysRevD.92.085032} {\bibfield  {journal} {\bibinfo
  {journal} {Phys. Rev.}\ }\textbf {\bibinfo {volume} {D92}},\ \bibinfo {pages}
  {085032} (\bibinfo {year} {2015}{\natexlab{a}})}\BibitemShut {NoStop}%
%%CITATION = PHRVA,D92,085032;%%
\bibitem [{\citenamefont {Gr\"utzmann}\ and\ \citenamefont
  {Strobl}(2015)}]{Gruetzmann-Strobl}%
  \BibitemOpen
  \bibfield  {author} {\bibinfo {author} {\bibfnamefont {M.}~\bibnamefont
  {Gr\"utzmann}}\ and\ \bibinfo {author} {\bibfnamefont {T.}~\bibnamefont
  {Strobl}},\ }\href {\doibase 10.1142/S0219887815500097} {\bibfield  {journal}
  {\bibinfo  {journal} {International Journal of Geom. Methods in Modern
  Physics}\ }\textbf {\bibinfo {volume} {12}},\ \bibinfo {pages} {1550009}
  (\bibinfo {year} {2015})}\BibitemShut {NoStop}%
\bibitem [{\citenamefont {Kotov}\ and\ \citenamefont
  {Strobl}(2015{\natexlab{b}})}]{KS07}%
  \BibitemOpen
  \bibfield  {author} {\bibinfo {author} {\bibfnamefont {A.}~\bibnamefont
  {Kotov}}\ and\ \bibinfo {author} {\bibfnamefont {T.}~\bibnamefont {Strobl}},\
  }\href {\doibase 10.1142/S0219887815500061} {\bibfield  {journal} {\bibinfo
  {journal} {International Journal of Geometric Methods in Modern Physics}\
  }\textbf {\bibinfo {volume} {12}},\ \bibinfo {pages} {1550006} (\bibinfo
  {year} {2015}{\natexlab{b}})},\ \Eprint {http://arxiv.org/abs/0711.4106}
  {arXiv:0711.4106 [math.DG]} \BibitemShut {NoStop}%
\bibitem [{\citenamefont {Lavau}\ \emph {et~al.}(2014)\citenamefont {Lavau},
  \citenamefont {Samtleben},\ and\ \citenamefont {Strobl}}]{Lavau}%
  \BibitemOpen
  \bibfield  {author} {\bibinfo {author} {\bibfnamefont {S.}~\bibnamefont
  {Lavau}}, \bibinfo {author} {\bibfnamefont {H.}~\bibnamefont {Samtleben}}, \
  and\ \bibinfo {author} {\bibfnamefont {T.}~\bibnamefont {Strobl}},\ }\href
  {\doibase 10.1016/j.geomphys.2014.10.006} {\bibfield  {journal} {\bibinfo
  {journal} {J. Geom. Phys.}\ }\textbf {\bibinfo {volume} {86}},\ \bibinfo
  {pages} {497} (\bibinfo {year} {2014})},\ \Eprint
  {http://arxiv.org/abs/1403.7114} {arXiv:1403.7114 [math-ph]} \BibitemShut
  {NoStop}%
%%CITATION = ARXIV:1403.7114;%%
\bibitem [{\citenamefont {Baez}\ and\ \citenamefont {Crans}(2003)}]{Baez03vi}%
  \BibitemOpen
  \bibfield  {author} {\bibinfo {author} {\bibfnamefont {J.~C.}\ \bibnamefont
  {Baez}}\ and\ \bibinfo {author} {\bibfnamefont {A.}~\bibnamefont {Crans}},\
  }\href@noop {} {\ ,\ \bibinfo {pages} {47} (\bibinfo {year} {2003})},\
  \Eprint {http://arxiv.org/abs/math.QA/0307263} {math.QA/0307263} \BibitemShut
  {NoStop}%
\bibitem [{\citenamefont {Lada}\ and\ \citenamefont
  {Stasheff}(1993)}]{Lieinfinityalg}%
  \BibitemOpen
  \bibfield  {author} {\bibinfo {author} {\bibfnamefont {T.}~\bibnamefont
  {Lada}}\ and\ \bibinfo {author} {\bibfnamefont {J.}~\bibnamefont
  {Stasheff}},\ }\href {\doibase 10.1007/BF00671791} {\bibfield  {journal}
  {\bibinfo  {journal} {Int. J. Theor. Phys.}\ }\textbf {\bibinfo {volume}
  {32}},\ \bibinfo {pages} {1087} (\bibinfo {year} {1993})},\ \Eprint
  {http://arxiv.org/abs/hep-th/9209099} {arXiv:hep-th/9209099 [hep-th]}
  \BibitemShut {NoStop}%
%%CITATION = HEP-TH/9209099;%%
\bibitem [{\citenamefont {Schwarz}(1993)}]{Schwarz}%
  \BibitemOpen
  \bibfield  {author} {\bibinfo {author} {\bibfnamefont {A.~S.}\ \bibnamefont
  {Schwarz}},\ }\href {\doibase 10.1007/BF02097392} {\bibfield  {journal}
  {\bibinfo  {journal} {Commun.Math.Phys.}\ }\textbf {\bibinfo {volume}
  {155}},\ \bibinfo {pages} {249} (\bibinfo {year} {1993})},\ \Eprint
  {http://arxiv.org/abs/hep-th/9205088} {hep-th/9205088 [hep-th]} \BibitemShut
  {NoStop}%
\bibitem [{\citenamefont {Alexandrov}\ \emph {et~al.}(1997)\citenamefont
  {Alexandrov}, \citenamefont {Kontsevich}, \citenamefont {Schwarz},\ and\
  \citenamefont {Zaboronsky}}]{AKSZ}%
  \BibitemOpen
  \bibfield  {author} {\bibinfo {author} {\bibfnamefont {M.}~\bibnamefont
  {Alexandrov}}, \bibinfo {author} {\bibfnamefont {M.}~\bibnamefont
  {Kontsevich}}, \bibinfo {author} {\bibfnamefont {A.}~\bibnamefont {Schwarz}},
  \ and\ \bibinfo {author} {\bibfnamefont {O.}~\bibnamefont {Zaboronsky}},\
  }\href@noop {} {\bibfield  {journal} {\bibinfo  {journal} {Int. J. Mod.
  Phys.}\ }\textbf {\bibinfo {volume} {A12}},\ \bibinfo {pages} {1405}
  (\bibinfo {year} {1997})},\ \Eprint {http://arxiv.org/abs/hep-th/9502010}
  {hep-th/9502010} \BibitemShut {NoStop}%
%%CITATION = HEP-TH 9502010;%%
\bibitem [{Note1()}]{Note1}%
  \BibitemOpen
  \bibinfo {note} {The basis $F^\alpha $ of generators of ${\protect \cal {}I}$
  is uniquely singled out by the property that the only term containing a de
  Rham differential is the first one, $\protect \mathrm {d}{}A^\alpha $.
  However, for squaring inside an action functional, as needed in a kinetic
  term for $A^\alpha $, such a choice of field strengths will turn out to be
  too restrictive.}\BibitemShut {Stop}%
\bibitem [{\citenamefont {Strobl}(2004)}]{LAYM}%
  \BibitemOpen
  \bibfield  {author} {\bibinfo {author} {\bibfnamefont {T.}~\bibnamefont
  {Strobl}},\ }\href {\doibase 10.1103/PhysRevLett.93.211601} {\bibfield
  {journal} {\bibinfo  {journal} {Phys. Rev. Lett.}\ }\textbf {\bibinfo
  {volume} {93}},\ \bibinfo {pages} {211601} (\bibinfo {year} {2004})},\
  \Eprint {http://arxiv.org/abs/hep-th/0406215} {arXiv:hep-th/0406215 [hep-th]}
  \BibitemShut {NoStop}%
%%CITATION = HEP-TH/0406215;%%
\bibitem [{\citenamefont {Henneaux}\ and\ \citenamefont
  {Teitelboim}(1992)}]{Henneaux-Teitelboim}%
  \BibitemOpen
  \bibfield  {author} {\bibinfo {author} {\bibfnamefont {M.}~\bibnamefont
  {Henneaux}}\ and\ \bibinfo {author} {\bibfnamefont {C.}~\bibnamefont
  {Teitelboim}},\ }\href@noop {} {\emph {\bibinfo {title} {Quantization of
  {G}auge {S}ystems}}}\ (\bibinfo  {publisher} {Princeton University Press},\
  \bibinfo {address} {Kassel},\ \bibinfo {year} {1992})\BibitemShut {NoStop}%
\bibitem [{Note2()}]{Note2}%
  \BibitemOpen
  \bibinfo {note} {Note that in \cite {Baez02jn} the important question of
  gauge invariance was ignored and the functional provided there was in fact
  not invariant.}\BibitemShut {Stop}%
\bibitem [{Note3()}]{Note3}%
  \BibitemOpen
  \bibinfo {note} {Most dramatic is the simplification in \protect \textup
  {\hbox {\mathsurround \z@ \protect \normalfont (\ignorespaces \ref
  {gamma}\unskip \@@italiccorr )}}. As seen best from \protect \textup {\hbox
  {\mathsurround \z@ \protect \normalfont (\ignorespaces \ref {two}\unskip
  \@@italiccorr )}}, where the second line vanishes completely after
  anti-symmetrization in the three indices, one finds $\gamma
  (v_1,v_2,v_3)=\protect \frac {1}{3}([v_1,v_2]_\protect \mathbb {V}\circ v_3 +
  cycl)$.}\BibitemShut {Stop}%
\bibitem [{Note4()}]{Note4}%
  \BibitemOpen
  \bibinfo {note} {Identify for this $C^a_{bc}$ with $-f^a_{bc}$, $t^a_I$ with
  $-h^a_I$ and $\beta ^I_{ab}$ with the negative of the symmetric $d^I_{ab}$
  and put to zero the 3-form field together with $b_{Iab}$ and
  $g^{Ia}$.}\BibitemShut {Stop}%
\bibitem [{\citenamefont {{Roytenberg}}(2009)}]{Courant-Dorfman-algebras}%
  \BibitemOpen
  \bibfield  {author} {\bibinfo {author} {\bibfnamefont {D.}~\bibnamefont
  {{Roytenberg}}},\ }\href {\doibase 10.1007/s11005-009-0342-3} {\bibfield
  {journal} {\bibinfo  {journal} {Letters in Mathematical Physics}\ }\textbf
  {\bibinfo {volume} {90}},\ \bibinfo {pages} {311} (\bibinfo {year} {2009})},\
  \Eprint {http://arxiv.org/abs/0902.4862} {arXiv:0902.4862 [math.QA]}
  \BibitemShut {NoStop}%
\bibitem [{\citenamefont {{Liu}}\ \emph {et~al.}(1995)\citenamefont {{Liu}},
  \citenamefont {{Weinstein}},\ and\ \citenamefont
  {{Xu}}}]{Courant-algebroids}%
  \BibitemOpen
  \bibfield  {author} {\bibinfo {author} {\bibfnamefont {Z.-J.}\ \bibnamefont
  {{Liu}}}, \bibinfo {author} {\bibfnamefont {A.}~\bibnamefont {{Weinstein}}},
  \ and\ \bibinfo {author} {\bibfnamefont {P.}~\bibnamefont {{Xu}}},\ }in\
  \href@noop {} {\emph {\bibinfo {booktitle} {eprint arXiv:dg-ga/9508013}}}\
  (\bibinfo {year} {1995})\BibitemShut {NoStop}%
\bibitem [{\citenamefont {Severa}\ and\ \citenamefont
  {Weinstein}(2001)}]{Severa-Weinstein}%
  \BibitemOpen
  \bibfield  {author} {\bibinfo {author} {\bibfnamefont {P.}~\bibnamefont
  {Severa}}\ and\ \bibinfo {author} {\bibfnamefont {A.}~\bibnamefont
  {Weinstein}},\ }\bibfield  {booktitle} {\emph {\bibinfo {booktitle}
  {{Noncommutative geometry and string theory. Proceedings, International
  Workshop, Yokohama, Japan, March 16-22, 2001}}},\ }\href {\doibase
  10.1143/PTPS.144.145} {\bibfield  {journal} {\bibinfo  {journal} {Prog.
  Theor. Phys. Suppl.}\ }\textbf {\bibinfo {volume} {144}},\ \bibinfo {pages}
  {145} (\bibinfo {year} {2001})},\ \Eprint {http://arxiv.org/abs/math/0107133}
  {arXiv:math/0107133 [math-sg]} \BibitemShut {NoStop}%
%%CITATION = MATH/0107133;%%
\bibitem [{\citenamefont {{Alekseev}}\ and\ \citenamefont
  {{Strobl}}(2005)}]{Alekseev-Strobl}%
  \BibitemOpen
  \bibfield  {author} {\bibinfo {author} {\bibfnamefont {A.}~\bibnamefont
  {{Alekseev}}}\ and\ \bibinfo {author} {\bibfnamefont {T.}~\bibnamefont
  {{Strobl}}},\ }\href {\doibase 10.1088/1126-6708/2005/03/035} {\bibfield
  {journal} {\bibinfo  {journal} {Journal of High Energy Physics}\ }\textbf
  {\bibinfo {volume} {3}},\ \bibinfo {eid} {035} (\bibinfo {year} {2005})},\
  \Eprint {http://arxiv.org/abs/hep-th/0410183} {hep-th/0410183} \BibitemShut
  {NoStop}%
\bibitem [{\citenamefont {{Roytenberg}}\ and\ \citenamefont
  {{Weinstein}}(1998)}]{Roytenberg-Weinstein}%
  \BibitemOpen
  \bibfield  {author} {\bibinfo {author} {\bibfnamefont {D.}~\bibnamefont
  {{Roytenberg}}}\ and\ \bibinfo {author} {\bibfnamefont {A.}~\bibnamefont
  {{Weinstein}}},\ }\href@noop {} {\bibfield  {journal} {\bibinfo  {journal}
  {Lett. Math. Phys.}\ }\textbf {\bibinfo {volume} {46}},\ \bibinfo {pages}
  {81} (\bibinfo {year} {1998})},\ \Eprint {http://arxiv.org/abs/math/9802118}
  {math/9802118} \BibitemShut {NoStop}%
\bibitem [{Note5()}]{Note5}%
  \BibitemOpen
  \bibinfo {note} {More details on this will be provided in \cite
  {Alexei-Henning-Strobl}.}\BibitemShut {Stop}%
\bibitem [{\citenamefont {Kotov}\ \emph {et~al.}(tion)\citenamefont {Kotov},
  \citenamefont {Samtleben},\ and\ \citenamefont
  {Strobl}}]{Alexei-Henning-Strobl}%
  \BibitemOpen
  \bibfield  {author} {\bibinfo {author} {\bibfnamefont {A.}~\bibnamefont
  {Kotov}}, \bibinfo {author} {\bibfnamefont {H.}~\bibnamefont {Samtleben}}, \
  and\ \bibinfo {author} {\bibfnamefont {T.}~\bibnamefont {Strobl}},\
  }\href@noop {} {} (\bibinfo {year} {in preparation})\BibitemShut {NoStop}%
\bibitem [{\citenamefont {de~Wit}\ \emph {et~al.}(2005)\citenamefont {de~Wit},
  \citenamefont {Samtleben},\ and\ \citenamefont {Trigiante}}]{tensor1}%
  \BibitemOpen
  \bibfield  {author} {\bibinfo {author} {\bibfnamefont {B.}~\bibnamefont
  {de~Wit}}, \bibinfo {author} {\bibfnamefont {H.}~\bibnamefont {Samtleben}}, \
  and\ \bibinfo {author} {\bibfnamefont {M.}~\bibnamefont {Trigiante}},\ }\href
  {\doibase 10.1016/j.nuclphysb.2005.03.032} {\bibfield  {journal} {\bibinfo
  {journal} {Nucl. Phys.}\ }\textbf {\bibinfo {volume} {B716}},\ \bibinfo
  {pages} {215} (\bibinfo {year} {2005})},\ \Eprint
  {http://arxiv.org/abs/hep-th/0412173} {arXiv:hep-th/0412173 [hep-th]}
  \BibitemShut {NoStop}%
%%CITATION = HEP-TH/0412173;%%
\bibitem [{\citenamefont {de~Wit}\ \emph {et~al.}(2008)\citenamefont {de~Wit},
  \citenamefont {Nicolai},\ and\ \citenamefont {Samtleben}}]{tensor2}%
  \BibitemOpen
  \bibfield  {author} {\bibinfo {author} {\bibfnamefont {B.}~\bibnamefont
  {de~Wit}}, \bibinfo {author} {\bibfnamefont {H.}~\bibnamefont {Nicolai}}, \
  and\ \bibinfo {author} {\bibfnamefont {H.}~\bibnamefont {Samtleben}},\ }\href
  {\doibase 10.1088/1126-6708/2008/02/044} {\bibfield  {journal} {\bibinfo
  {journal} {JHEP}\ }\textbf {\bibinfo {volume} {02}},\ \bibinfo {pages} {044}
  (\bibinfo {year} {2008})},\ \Eprint {http://arxiv.org/abs/0801.1294}
  {arXiv:0801.1294 [hep-th]} \BibitemShut {NoStop}%
%%CITATION = ARXIV:0801.1294;%%
\bibitem [{\citenamefont {de~Wit}\ and\ \citenamefont
  {Samtleben}(2008)}]{tensor3}%
  \BibitemOpen
  \bibfield  {author} {\bibinfo {author} {\bibfnamefont {B.}~\bibnamefont
  {de~Wit}}\ and\ \bibinfo {author} {\bibfnamefont {H.}~\bibnamefont
  {Samtleben}},\ }\href {\doibase 10.1088/1126-6708/2008/08/015} {\bibfield
  {journal} {\bibinfo  {journal} {JHEP}\ }\textbf {\bibinfo {volume} {08}},\
  \bibinfo {pages} {015} (\bibinfo {year} {2008})},\ \Eprint
  {http://arxiv.org/abs/0805.4767} {arXiv:0805.4767 [hep-th]} \BibitemShut
  {NoStop}%
%%CITATION = ARXIV:0805.4767;%%
\end{thebibliography}
\end{document}